\documentclass[prl,twocolumn,showpacs,amsmath,superscriptaddress]{revtex4}

\usepackage{epsfig}
\usepackage{graphicx}
\usepackage{amsmath}
\usepackage{amssymb}
\usepackage{subfigure}

\begin{document}

\title{Non-Universal Extinction Transition for Boundary Active Site}

\author{S. Burov}
\affiliation{Department of Physics, Bar Ilan University, Ramat-Gan 52900,
Israel}
\author{D. A.  Kessler}
\affiliation{Department of Physics, Bar Ilan University, Ramat-Gan 52900,
Israel}


\begin{abstract}
We present a generalized model of a diffusion-reaction system where the reaction
occurs only on the boundary.  This model reduces to that of Barato and
Hinrichsen when the occupancy of the boundary site is restricted to zero or one.
 In the limit when there is no restriction on the occupancy of the boundary
site, the model reduces to an  age dependent Galton-Watson branching process and
admits an analytic solution.   The model 
displays a boundary-induced phase transition into an absorbing state 
with rational critical exponents and exhibits aging at criticality below a
certain fractal dimension of the diffusion process. Surprisingly the behavior
in the 
critical regime for intermediate occupancy restriction $N$ varies with $N$.
In fact, by varying the lifetime of the active boundary particle or the
diffusion coefficient in the bulk, the critical exponents can be continuously
modified. \end{abstract}

\maketitle
Nonequilibrium phase transitions can differ significantly from their equilibrium cousins, and are a subject of continuing interest.
 Recently, 
Barato and Hinrichsen~\cite{Hinrichsen01,Hinrichsen02} (BH) studied a reaction-diffusion model apparently exhibiting a new universality class  of nonequilibrium phase transitions,  boundary induced phase 
transitions.  The model they studied was a variant of a model introduced earlier by
Deloubri\`ere and van Wijland~\cite{first}, who however did not discover the novel scaling behavior.
In this letter we generalize the model and solve it analytically in a certain limit. 
Our model exhibits a phase transition and at criticality we find the presence 
 of aging~\cite{Glasses,Phleming} , a nonequilibrium property observed in such
diverse systems 
as spin-glasses~\cite{Glasses}, gels~\cite{Weitz} and
turbulence~\cite{Grigolini02}. 
%as aging[ADD REFERNCES] is observed, 
Furthermore a critical fractional dimension for the diffusion process
is naturally obtained in the context of our model. Interestingly we find that the critical 
exponents of the model vary continuously with the parameters of the model. 
We also note that our toy model is  interesting in the context of 
catalytic reactions and biological situations when the reaction occurs on a 
specific membrane, such as a boundary of a cell.

Our model is defined as follows: As in the BH model, a particle starting on the boundary follows 
a birth-death process.  It produces an offspring (next to the boundary site) with probability $p$ 
($A\displaystyle{\xrightarrow{p}} A + O$) and dies with probability $1-p$ 
 ($A\displaystyle{\xrightarrow{1-p}} \varnothing$). The particle $A$ continues 
to reproduce until it dies. The offspring $O$ diffuse freely in the bulk since their birth and up to 
the time they reach the boundary, and then they start to reproduce again according 
to the same death-birth process. By varying the rate of offspring production, 
one expects to reach extinction for $p\to 0$ (absorbing state) and a growing 
or stable population for $p\to1$, intuitively one expects a transition between those 
two states as $p$ grows from $0$ to $1$. In the BH model, there is a constraint that only one particle  can exist  on the boundary site at any give time. Any other particle trying to enter the boundary site dies immediately. We generalize this to allow up to $N$ particles to coexist simultaneously on the boundary.  As shown by
BH,   the role of the bulk 
diffusion is just to generate a probability distribution  $\psi(t)$ for the arrival of the newly born 
particle to the boundary site, where it can, if the boundary site is not fully occupied, reproduce some number of times, setting off  new processes and then die. 
Introducing constraints on the occupancy of the bulk sites has no effect on the global dynamics, due to the indistinguishability of different particles~\cite{Levitt}. 
By totally relaxing  the constraint on the boundary occupancy, i.e., $N\to\infty$, we are capable of completely solving the model 
since technically the unrestricted model is a random branching, or Galton-Watson, process.

In the unrestricted case we can make another simplification, since the time to die is governed 
by a short-range exponential distribution, we can, without doing any harm to the model, consider all births as happening simultaneously.  
In this view, the particle generates $k$ children, governed by a geometrical distribution, $p_k =(1-p)p^k$, and then immediately dies.
The basic technique employed in the analysis is the use of the generating functional~\cite{Redner02} and parallels the solution of the standard age-dependent branching process~\cite{Grimmet}.
Let us denote by $Z(t,t+s)$ the number of particles that been observed on the boundary during 
the time interval $(t,t+s)$.
We define a generating function for the random variable $Z(t,t+s)$ as the average of $s^{Z(t,t+s)}$ over the 
distribution of $Z(t,t+s)$ 
\begin{equation}
G_{(t,t+s)}(s) = \sum_{Z(t,t+s)=0}^\infty\text{P}(Z(t,t+s)) s^{Z(t,t+s)}   ,
\label{def_generat}
\end{equation}
where $\text{P}(Z(t,t+s))$ denotes the probability to observe $Z(t,t+s)$ particles on the boundary in the 
specified time interval. The generating function for the number of offsprings is $G_O(s)=\langle s^k \rangle=(1-p)/(1-ps)$. 
We set the initial conditions such that at $t=0$ a single particle was injected into the bulk 
and it has the probability distribution $\psi(t)$ to return to the boundary at time $t=t_R$. By conditioning on the outcome 
of the returning time for the first particle~\cite{Grimmet}, we
obtain~\cite{Burov} 
the equation for $G_{(t,t+s)}(s)$ 
\begin{equation}
\begin{array}{l}
\displaystyle{ G_{(t,t+s)}(s) = \int_0^t G_O(G_{(t-u,t-u+s)}(s))\psi(u)du }
\\
\displaystyle{
+s \int_t^{t+s} G_O(G_{(0,t-u+s)}(s))\psi(u)du
+\int_{t+s}^\infty \psi(u)du}.
\end{array}
\label{generatin1}
\end{equation}
In the limit of $s\to0$ the generating function goes to $P(Z(t,t+s)=0):=P_E(t,s)$, i.e. 
the probability that not a single particle appears on the boundary at the mentioned time 
interval, taking this limit in Eq.~(\ref{generatin1}) we obtain
\begin{equation}
\begin{array}{l}
 P_E(t,s)=
\\
\displaystyle{\int_0^t \frac{1-p}{1-pP_E(t-u,s)} \psi(u)du + \int_{t+s}^\infty \psi(u)du},
\end{array}
\label{general_fun}
\end{equation}
where we have now used the explicit from of $G_O(s)$. Eq~(\ref{general_fun}), a non-linear Volterra equation of the second 
kind,  is our main equation for the unrestricted case of the model and 
it describes the time evolution of a two-time quantity. We haven't assumed anything as to the form of 
$\psi(t)$ and so Eq.~(\ref{general_fun}) is quite general.  The long-time behavior of $P_E$ is, as we shall see, governed entirely by the long time behavior of $\psi(t)$.
For normal diffusion in the bulk, this long-time behavior is $\psi_t \sim \psi_\infty t^{-(1+\beta)}$, with $\beta=1/2$, $\psi_\infty = 1/\sqrt{4\pi D}$, where 
$D$ is the diffusion constant (defined as usual).  We also consider the more general case of $0<\beta<1$, and in such case the constant $\psi_\infty$ 
is used for the normalization of $\psi_t$. The case of $0<\beta<1$ can be achieved if the diffusion in the bulk is anomalous~\cite{Metzler};
e.g., subdiffusion for $0<\beta<1/2$ described by models such as the continuous
time random walk (CTRW)~\cite{Bouchaud,Metzler}. Herein we study $P_E$
exclusively; other quantities such as the mean number of particles can also be
obtained from Eq.~(\ref{generatin1})~\cite{Burov}.

We will now explore the behavior of $P_E(t,s)$ in the asymptotic limit of large $t$
for different values of the parameters $p$ and $\beta$. Inspired by numerical solutions of Eq. ~(\ref{general_fun}), we adopt the  ansatz 
\begin{equation}
P_E(t,s)=P_E^\infty-A(s)t^{-\alpha}
\qquad t\to\infty 
\label{ansatz}
\end{equation}
In order to 
obtain a solution of Eq.~(\ref{general_fun}) we substitute the asymptotic form in Eq.~(\ref{ansatz}) into both 
sides of Eq.~(\ref{general_fun}), perform an expansion in inverse powers of $t$ and compare the coefficients in front of the appropriate leading terms. 
We leave the exact technical details for a longer publication~\cite{Burov} and now provide only the final results.

So doing, in the limit $t\to\infty$, $s$ fixed,
we obtain 
\begin{equation}
P_E^\infty = \left\{
\begin{array}{ll}
\frac{1-p}{p} & p\geq1/2
\\
1	& p\leq1/2
\end{array}
\right.
\label{p_e_inf01}
\end{equation}
which clearly points out the existence of a critical $p=p_c=1/2$. The location of the critical point at $p=1/2$ is due to the fact that at this value, each particle on the boundary produces exactly one offspring. Below this point, the number of particles decreases exponentially with the number of past boundary particles, and above it it grows exponentially.  The behavior of $A(s)$ and $\alpha$ for the 
off-critical state, $p\neq p_c$, is given by 
\begin{equation}
A(s)=  \left\{
\begin{array}{ll}
\frac{1-p}{1-2p}\frac{\psi_0}{\beta}s & p<1/2
\\
-1 & p>1/2
\end{array}
\right.,
\label{a_delta01}
\end{equation}
\begin{equation}
\alpha = \left\{
\begin{array}{ll}
 1+\beta & p<1/2
\\
\beta & p>1/2
\end{array}
\right..
\label{alpha01}
\end{equation}
The presence of the phase transition as we approach $p=1/2$ from below is
clearly manifest in the divergence of the coefficient $A$.  The approach from
above is not obvious from the large $t$ behavior.  What happens is that for
$p\gtrsim 1/2$, $P_E$ first rises toward unity, as happens below the transition.
 However, at very large $t$, the behavior crosses over toward the power-law
decay toward $P_E^\infty$.  The details of this crossover will be presented in
our longer presentation~\cite{Burov}.

 In the critical state, $p=p_c$ Eq.~(\ref{p_e_inf01}) remains valid and so 
$P_E^\infty = 1$ while the behavior of $A(s)$ and $\alpha$  shows a transition  as a function of 
$\beta$.  For $\alpha$, we obtain
\begin{equation}
\alpha = \left\{
\begin{array}{ll}
 1-\beta & \beta\leq1/2
\\
\beta & \beta\geq1/2
\end{array}
\right.,
\label{alpha02}
\end{equation} 
We can exhibit an analytic expression for $A(s)$ at $p_c$ only for $\beta>1/2$:
\begin{equation}
A(s)= -\frac{\pi\csc(\beta\pi)\Gamma(1-b)}{\Gamma(1-2 \beta) \Gamma(1+\beta)}\qquad (\beta>1/2)
\label{a_delta02}
\end{equation}
For $\beta<1/2$, $A(s)$ has to be calculated numerically in general.  However, in the limit of large $s$, we have
\begin{equation}
A(s) \propto s^{1-\beta}\qquad (\beta<1/2).
\label{a_delat02}
\end{equation}
The existence of a special $\beta$ for the behavior at criticality is very non-trivial and 
it is especially interesting that the critical $\beta_c$ is equal to $1/2$, i.e the normal diffusion case. 
We can treat this critical $\beta_c$ as a critical fractal dimension since  
$2\beta$ is just the fractal dimension of the diffusion. For the special case of 
$\beta=1/2$ our ansatz, i.e. Eq.~(\ref{ansatz}), does not work and one needs to treat this case specially~\cite{Burov}; the result  is 
\begin{equation}
 1-P_E(t,s) \propto \frac{1}{\log(s\big/t)} \left(s \big/t\right)^{1/2}\qquad(\beta=1/2). 
\label{a_delta03}
\end{equation}
The logarithmic corrections in the behavior support our claims as to the
critical nature of  $\beta=1/2$. 
For a two time quantity like the survival probability $P_S(t,s)=1-P_E(t,s)$, 
one usually expects for a stationary process a  dependence only on the time difference $s$; when the process is non-stationary this is generally
not true. When the time dependence is that of a ratio of the two times this is usually 
defined as aging, as in our case $P_S(t,s)\propto (s/t)^{1-\beta}$. 
The aging behavior for $\beta\leq 1/2$ at criticality 
is a signature of the nonequlibrium phase transition properties, which have been studied extensively in the context 
of DP and Contact Process~\cite{Hinrichsen03,Gambassi,Phleming}, and the nonstationarity 
of the process usually obtained in glassy dynamics~\cite{Glasses}. We must note that the obtained result 
where the probability to observe at least one particle in time interval $s$ is 
proportional to $s^{1-\beta}$ ($\beta<1/2$) is unexpected in light of the fact that for the off-critical 
state ($p<p_c$), this probability is proportional to $s$.

Now we treat the limit $s\gg t \gg 1$, where we take first $s\to\infty$. In this limit 
we can neglect the second term on the right hand side of Eq.~(\ref{general_fun}) and define $P_E(t,\infty)=P_E(t)$ simply 
as the extinction probability of the process. We are now dealing with a single-time quantity and so 
no aging behavior could be obtained. For the solution we use the same ansatz as in Eq.~(\ref{ansatz}), writing
 $A(\infty)=A$. The equation for $P_E^\infty$ remains the same as 
Eq.~(\ref{p_e_inf01}) and one again obtains the same critical value for $p$, $p_c=1/2$.
In the off-critical state the results for $\alpha$ and $A$ are
\begin{equation}
 \alpha = \beta,
\label{alpha03}
\end{equation}
and 
\begin{equation}
 A(s)=  \left\{
\begin{array}{ll}
\frac{1-p}{1-2p}\frac{\psi_0}{\beta} & p<1/2
\\
\frac{1-p}{1-2p}\frac{\psi_0}{\beta} & p>1/2
\end{array}
\right. .
\label{A_01}
\end{equation}
The transition in this limit is different then in the previously discussed limit of $t\gg s$, as $A$ diverges 
similarly near $p_c$  and thus describes the crossover time scale on which one would observe  critical behavior. For the critical state $p=p_c$ we obtain 
\begin{equation}
P_E(t)\sim 1 - \left(\frac{\psi_0}{\beta} \right)^{\beta/2} t^{-\beta/2}.
\label{critical2}
\end{equation}
Thus for the survival probability of the process, $P_S(t)=1-P_E(t)$, we obtain the power-law 
behavior $P_S(t)\sim t^{-\delta}$ with $\delta=\beta/2$. The special properties of $\beta=1/2$ are not observed
in the limit of $s \gg t$, and as already been mentioned no aging behavior 
could be observed for a one-time quantity. The noncomutativity of the limits $t\to\infty$ - $s\to\infty$ is similar 
in some sense to the non commutativity of the same limits present for correlation function 
behavior in glassy systems~\cite{Glasses} and leads to such non-equilibrium property as ergodicity breaking~\cite{Bouchaud1,Bel}

Thus, in the case of normal diffusion, i.e. $\beta=1/2$, for our modified model
with unrestricted occupation of the origin, we have a phase transition at
$p=1/2$ with a critical exponent for the survival probability of $\delta =
1/4$, observed also in numerical simulations~\cite{Hinrichsen02} .  This is in
contrast with the original model, with a larger critical $p$, but more
importantly, a survival exponent of $\delta\approx 0.17$. The change in the
phase transition point is clear, since in the original $N=1$ model, some
children die upon their attempted return to the origin and so the critical $p$
has to be larger than 1/2 to compensate.  The change in exponent is not
unexpected.  For directed percolation, the critical exponents in a model of
unrestricted occupancy are the mean-field exponents, which differ from the
exponents in the restricted occupancy version\cite{DPN}.

In the case of directed percolation, for any finite $N$ the critical exponents are those of the $N=1$ model.  For large $N$, there is a crossover in the scaling between the mean-field scaling of the $N=\infty$ model and the directed percolation exponents~\cite{DPN}.  Based on this analogy, one would expect a similar behavior in the present case, with the critical behavior for any finite $N$ being identical to the $N=1$ model.  

\begin{figure}
\center{
\includegraphics[width=9.5cm]{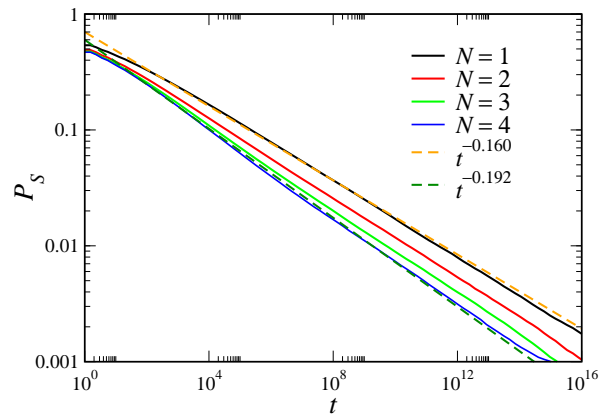}}
\caption{The survival probability vs. time for $N=1$, 2, 3, and 4, together with power-law fits for the cases $N=1$ and $N=4$.}
\label{fig1}
\end{figure}

\begin{figure}
\center{\includegraphics[width=9.5cm]{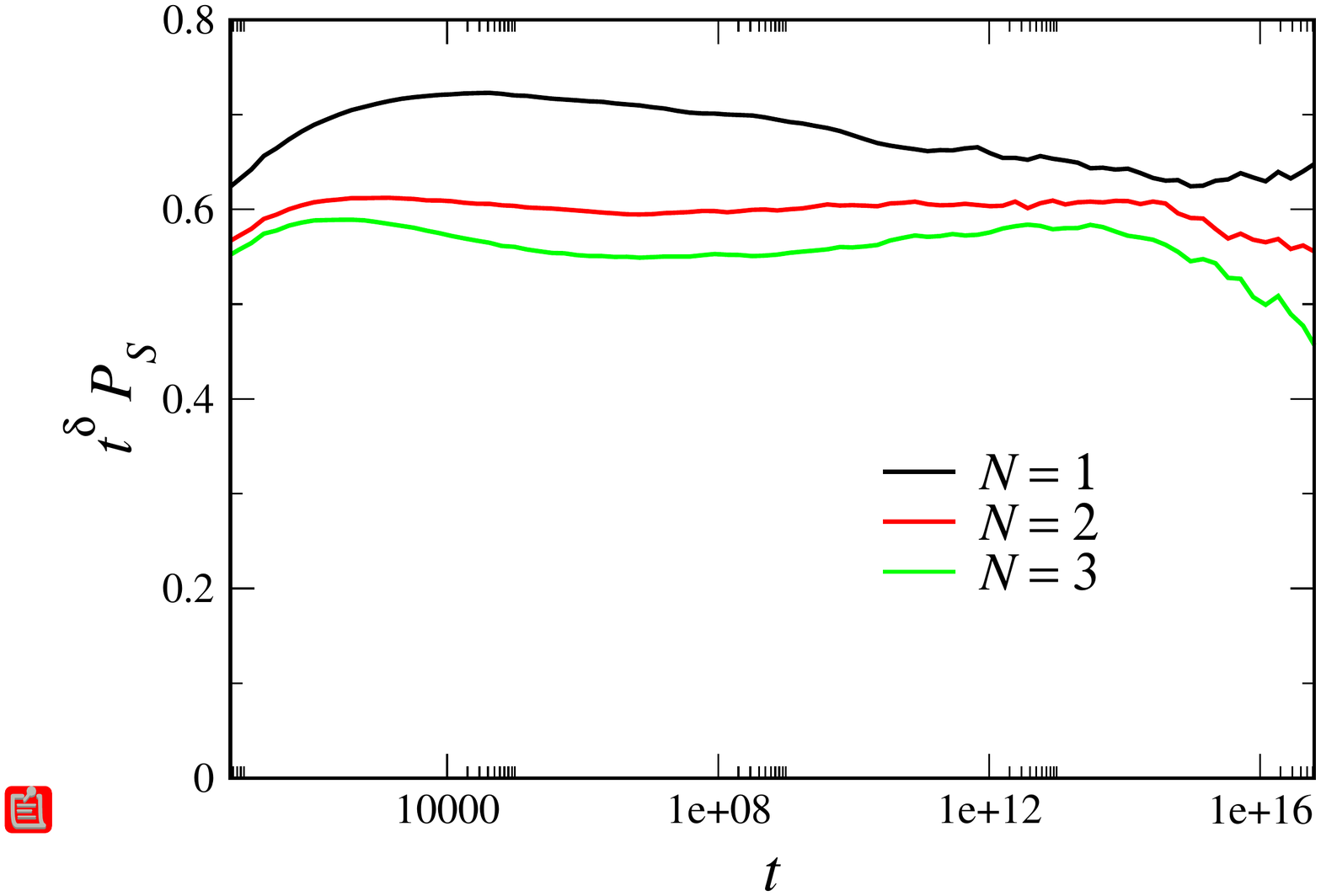}}
\caption{The survival probability, $P_S^N$, multiplied by $e^{\delta_N t}$ as a function of time for $N=1$, $2$ and $3$.  The values of $\delta_N$ are as
noted in the text.}
\label{fig2}
\end{figure}

To test this hypothesis, we have measured the survival exponent $\delta$ at
criticality for various $N$.  The results are shown in Fig. \ref{fig1}.  We see
that, contrary to our naive expectation, there is a different exponent for each
$N$.  The measured best fit exponents are, for example, $\delta=0.160$ for
$N=1$, $\delta=0.171$ for $N=2$, $\delta=0.180$ for $N=3$ and $\delta=0.192$
for $N=4$.
There is no sign of any crossover behavior.  One test for this is shown in Fig.
\ref{fig2}, where we show 
$P_S(t) t^\delta$, in a linear($y$)-log($x$) scale.  We see that the graphs all exhibit
the same characteristic oscillation in $\ln(x)$~\cite{sournette} that makes accurate determination of an exponent so difficult, but they show no secular trend.

\begin{figure}
\begin{center}
\includegraphics[width=9.5cm]{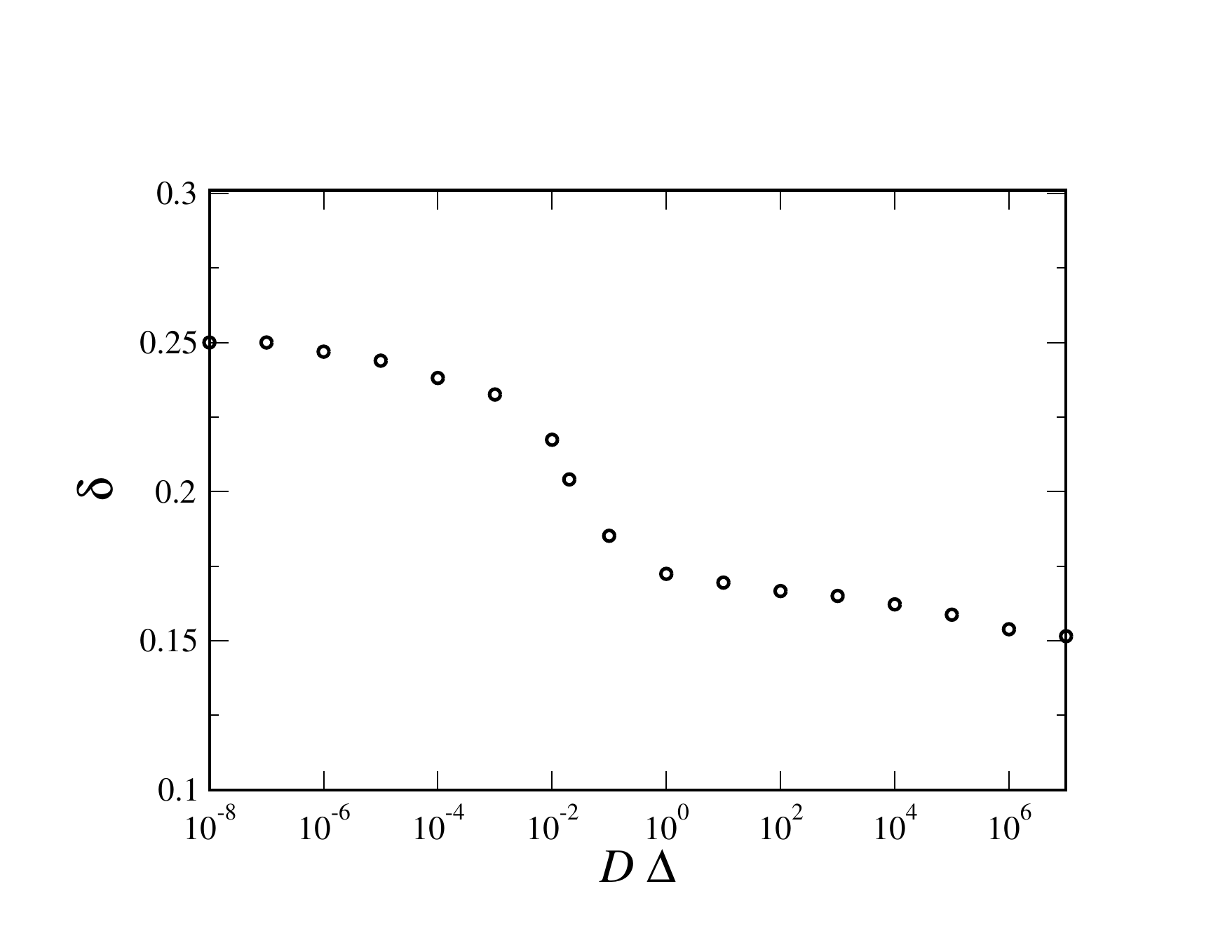}
\caption{The measured survival exponent $\delta$ as a function of $\Delta$, the
lifetime of the particle on the active site after it gives birth. The
simulations were performed for measurement time $t\sim10^{16}$.}
\end{center}
\label{fig3}
\end{figure}

Varying $N$ is not the only way to interpolate between the two models and thereby vary $\delta$.
We can instead introduce a parameter $\Delta$ which controls the lifetime of a particle on the active site after reproduction (in the original model, $\Delta=1$).  As $\Delta$ decreases, the interference between different particles is reduced, and the theory approaches the noninterference, Galton-Watson, limit. We can also increase $\Delta$ beyond one, and make the interference effect stronger, thereby reducing the exponent below the original BH value. Actually the precise dimensionless parameter which controls the exponent $\delta$ is 
$\Delta/\psi_\infty^2 \propto D\Delta$, which presents a surprising dependence of the critical exponent $\delta$ on the 
diffusion coefficient ($D$) in the bulk.  This is demonstrated in
Fig.~\ref{fig3}, where the survival exponent is plotted as a function of
$\Delta$.  We see that as opposed to the discrete parameter $N$, now the
exponent varies continuously, decreasing with $D\Delta$. 

%\begin{figure}
%\centering
%\includegraphics[width=8cm,height =5cm,viewport=0 400 700 800]{output.ps}
%\caption{BLABLA.
%}
%\label{fig2}
%\end{figure}

Finally we note that the relevance of the limit for which we managed to solve the model analytically is 
much broader when one considers higher dimension systems. We have studied herein  a one dimensional 
system where the boundary is zero dimensional, while if one considers a $d$ dimensional system 
where the boundary is of dimension $d-1$, a line or a membrane, in such a case
even if $D\Delta$ does not  approach zero, the probability that two particles
would try to 
occupy the same site (since there an infinite number of sites on boundary) during the time $D \Delta$ is small and this would lead to a larger applicability of our analytical solution. Another type of 
system where our results could be applied is systems with anomalous diffusion behavior described 
by the Continuous Time Random-Walk model~\cite{Bouchaud}, for such systems the results 
stays valid even if the size of the system is finite as for a finite system with regular diffusion 
our results are modified for times long enough~\cite{Burov}.

In summary, we have  extended to finite active site occupancy $N$ and solved analytically the $N\to\infty$ limit of the Barato-Hinrichsen model, yielding the scaling exponents of the extinction transition in this limit.  We have also show that below a critical fractal dimension for the diffusion process in the bulk, there is aging behavior at the transition.  In addition, we have seen that the exponents are in fact not universal, varying with $N$ or the scaled lifetime of the particles on the active site.  This is of course very different from the situation that obtains in the superficially similar contact process, which exhibits universal Directed Percolation scaling.

\end{document}